\documentclass[twocolumn,aps,prl,floatfix,superscriptaddress]{revtex4-1}
\usepackage{amsmath,amssymb}
\usepackage{graphicx,float}
\usepackage{times,txfonts}
\usepackage{color}
\usepackage{multirow}
\usepackage{bbold}
\usepackage{color}
\usepackage{soul}
\usepackage{hyperref}
\usepackage{times,txfonts}
\usepackage{natbib}
\usepackage{blindtext}
\usepackage[export]{adjustbox}
\usepackage{mathrsfs}
\usepackage{textcomp}

\newcommand{\etal}{\textit{et~al.}}
\renewcommand{\epsilon}{\varepsilon}
\renewcommand{\phi}{\varphi}

\usepackage{sistyle}
\usepackage{soul}
\definecolor{lightblue}{RGB}{185,210,248}
\sethlcolor{lightblue}

\begin{document}
\title{Towards a holographic approach to spherical aberration correction in scanning transmission electron microscopy}
\author{Vincenzo~Grillo}
\affiliation{CNR-Istituto Nanoscienze, Centro S3, Via G Campi 213/a, I-41125 Modena, Italy}
\author{Amir~H.~Tavabi}
\affiliation{Ernst Ruska-Centre for Microscopy and Spectroscopy with Electrons, Forschungszentrum J\"ulich, 52425~J\"ulich, Germany}
\author{Emrah~Yucelen}
\affiliation{Thermo Fisher Company, PO Box 80066, 5600KA, Eindhoven, The Netherlands}
\author{Peng-Han~Lu}
\affiliation{Ernst Ruska-Centre for Microscopy and Spectroscopy with Electrons, Forschungszentrum J\"ulich, 52425~J\"ulich, Germany}
\author{Federico~Venturi}
\affiliation{CNR-Istituto Nanoscienze, Centro S3, Via G Campi 213/a, I-41125 Modena, Italy}
\affiliation{Dipartimento FIM Universit\'a di Modenae Reggio Emilia, Via G Campi 213/a, I-41125 Modena, Italy}
\author{Hugo~Larocque}
\affiliation{Department of Physics, University of Ottawa, 25 Templeton St., Ottawa, Ontario, K1N 6N5 Canada}
\author{Lei~Jin}
\affiliation{Ernst Ruska-Centre for Microscopy and Spectroscopy with Electrons, Forschungszentrum J\"ulich, 52425~J\"ulich, Germany}
\author{Aleksei~Savenko}
\affiliation{Thermo Fisher Company, PO Box 80066, 5600KA, Eindhoven, The Netherlands}
\author{Gian~Carlo~Gazzadi}
\affiliation{CNR-Istituto Nanoscienze, Centro S3, Via G Campi 213/a, I-41125 Modena, Italy}
\author{Roberto~Balboni}
\affiliation{CNR-IMM Bologna, Via P. Gobetti 101, 40129 Bologna, Italy}
\author{Stefano~Frabboni}
\affiliation{CNR-Istituto Nanoscienze, Centro S3, Via G Campi 213/a, I-41125 Modena, Italy}
\affiliation{Dipartimento FIM Universit\'a di Modenae Reggio Emilia, Via G Campi 213/a, I-41125 Modena, Italy}
\author{Peter~Tiemeijer}
\affiliation{Thermo Fisher Company, PO Box 80066, 5600KA, Eindhoven, The Netherlands}
\author{Rafal~E.~Dunin-Borkowski}
\affiliation{Ernst Ruska-Centre for Microscopy and Spectroscopy with Electrons, Forschungszentrum J\"ulich, 52425~J\"ulich, Germany}
\author{Ebrahim~Karimi}
\email{ekarimi@uottawa.ca}
\affiliation{Department of Physics, University of Ottawa, 25 Templeton St., Ottawa, Ontario, K1N 6N5 Canada}
\affiliation{Department of Physics, Institute for Advanced Studies in Basic Sciences, 45137-66731 Zanjan, Iran}
%
\begin{abstract}
Recent progress in phase modulation using nanofabricated electron holograms has demonstrated how the phase of an electron beam can be controlled. In this paper, we apply this concept to the correction of spherical aberration in a scanning transmission electron microscope and demonstrate an improvement in spatial resolution. Such a holographic approach to spherical aberration correction is advantageous for its simplicity and cost-effectiveness.
\end{abstract}
\pacs{Valid PACS appear here}
\maketitle

\section{Introduction}

The successful correction of spherical aberration in the late 1990s solved a long-standing limitation in transmission electron microscopy (TEM)~\cite{haider1998electron,krivanek1999towards,hawkes2009aberration}. This limitation, which afflicted the field since Scherzer discovered that the use of cylindrically symmetrical electrostatic and magnetic lenses leads to strictly positive spherical aberration~\cite{scherzer1936einige}, results from the impossibility of generating arbitrary fields in vacuum. Spherical aberration must be compensated by an opposite aberration, which is usually produced by a set of electric and magnetic multipoles, whose complexity rivals that of the adaptive optics of the Hubble telescope~\cite{allen1990hubble}. The primary complication of this approach is associated with the use of strong magnetic fields in coupled multipoles that must be highly stable and perfectly matched. As a consequence, there is still a need to develop alternative aberration correction concepts.

Recent progress in nanofabrication, in combination with inspiration drawn from methods in light optics, is resulting in the development of new innovative phase elements for electron microscopy~\cite{uchida2010generation,mcmorran2011electron,grillo2014highly}. Unprecedented control over the phase of an electron beam can now be achieved by using nanofabricated electron holograms, which consist of electron-transparent materials that are patterned to have controlled thickness modulations. Amplitude holograms can also be used to introduce phase variations in an electron beam by means of an initial amplitude modulation~\cite{verbeeck2010production}, although they are not as efficient as phase elements. Phase holograms, which are also known as kinoforms, can be classified into two categories. First, there are diffractive (or ``off-axis'') holograms~\cite{mcmorran2011electron,grillo2014highly}, which consist of diffraction gratings and control the phase of an electron beam through modulations in the spacing of the grating. Second, there are in-line holograms~\cite{uchida2010generation,shiloh2014sculpturing}, in which the phase that is imparted to the electron beam is directly proportional to the thickness profile of the material. Such devices were first used for the generation of electron vortex beams and subsequently in further applications that demonstrated their flexibility~\cite{voloch2013generation,grillo2014generation,grillo2017measuring}.
The ability to arbitrarily modulate the phase profile of an electron beam has led several groups to independently propose holographic methods for the correction of spherical aberration. For example, Linck \etal~\cite{linck2014aberration} and Grillo \etal~\cite{grillo2015structured,grilloToward2014} proposed a diffractive approach, but did not show how to remove the spurious diffracted beams that appear in the specimen plane of the microscope. Some of the co-authors of this paper initially used a blazed hologram~\cite{grillo2014highly,grilloToward2014} to transfer most of the intensity of the electron beam to a single spot.
Shiloh \etal~\cite{shiloh2016prospects} proposed using an in-line hologram for the correction of the main aberrations of an electron probe. Such an in-line approach has several advantages over its diffractive counterpart, in part due to the fact that the ray path is aligned with the propagation axis of the electron beam. However, it is challenging to implement in practice because the thickness of the slab must be controlled with nm precision over a large transverse area. In addition, regions where the phase wraps over $2\pi$ must also be accounted for by a sudden change in thickness. A failure to introduce the correct abrupt phase jump results in additional unwanted intensity components in the probe, even in cases where the overall phase correction is accurate. In contrast, an advantage of diffractive off-axis holograms over in-line holograms is that the selection of a single diffraction order automatically results in energy filtering of the electron beam that has passed through the hologram, as inelastically scattered electrons cannot be diffracted by a grating that typically has a periodicity of at least 50~nm.

Here, we demonstrate that a combination of an off-axis diffractive hologram and modified electron illumination optics can be used to produce an aberration corrected probe in a scanning TEM (STEM). The key point of our work is control over spherical aberration in a single isolated diffraction spot, opening the way to holographic control over aberrations.

\section{Materials and Methods}

\begin{figure}[h]
	\begin{center}
	\includegraphics[width=\columnwidth]{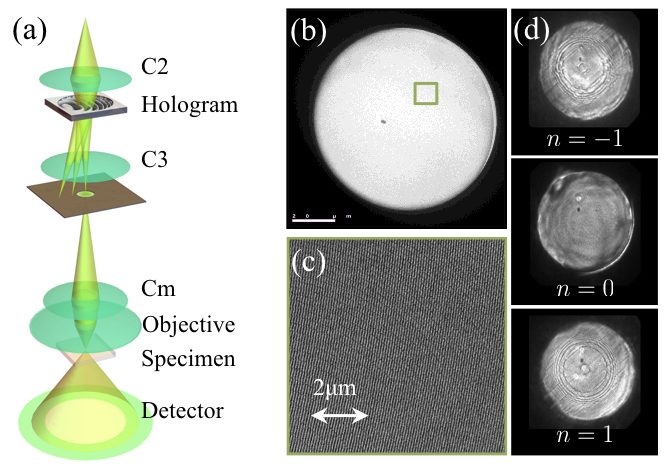}
	\caption[]{a)~Electron microscope configuration used to isolate a single holographically-corrected electron beam. b)~Scanning electron micrograph of a hologram that was placed in the C2 aperture plane. c)~Magnified image of b). d)~``Dark field'' images of the hologram recorded using individual diffraction orders. The images were obtained with the microscope in ``diffraction mode''.}
	\label{fig:fig1}
	\end{center}
\end{figure}

Fig.~\ref{fig:fig1} shows a schematic diagram of our setup, in which we use a nanofabricated hologram to generate a single spherical aberration corrected beam. The configuration is based on the typical illumination system of an FEI Titan TEM, with the convergence and defocus of the electron beam controlled independently by the coupled excitation of the two condenser lenses C2 and C3. The remaining lenses, including the objective lens and the condenser mini-lens Cm, are usually fixed. In the present configuration, we placed a nanofabricated hologram in the C2 condenser aperture plane. We then changed the excitation of the C3 lens, so that an intermediate image of the field emission gun (FEG) (\emph{i.e.}, the cross-over) was at the C3 aperture plane, while Cm was adjusted so that the probe was focused on the specimen. The convergence semi-angle could be adjusted by changing the C2 lens and refocusing with the C3 lens.
This modification was achieved by over-riding the standard working conditions of the microscope, which normally only allow for coupled action of the C2 and C3 condenser lenses over a limited range. In order to find the position of the crossover after the C2 lens, we searched for a configuration for which the excitation of C3 did not affect the beam. We found that a small defocus of C2 could be used to localize this crossover in the C3 aperture plane.
We were then able to select a single chosen beam that had been diffracted by the C2 hologram, thereby eliminating much of the diffuse scattering arising from it. Particular care was taken with the beam alignment to ensure that it was ``coma-free'' before reaching the aperture. A similar approach was recently proposed for the isolation of vortex beams~\cite{krivanek2014toward,pohl2016towards}.

In order to design a hologram that can correct for spherical aberration, we recall that the phase that is imparted by an electromagnetic lens onto a generic wavefront can be formulated in Fourier space as~\cite{kirkland2010advanced}
\begin{align}
	\begin{split}
	\label{eq:eq1}
	\chi = -\pi &\Delta f \lambda k^2 +\tfrac{\pi}{2}\lambda^3 C_s k^4 + \pi \text{A}_2 \lambda k^2\cos{(2(\theta-\theta_{2A}))} \\
	&+ \tfrac{2\pi}{3}\lambda^2\text{A}_3 k^3 \cos{(3(\theta-\theta_{3A}))}+\tfrac{2\pi}{3}\lambda^2 \text{B}_2 k^3 \cos{(\theta-\theta_B)}+ \dots~~,
	\end{split}
\end{align}
where $\lambda$ is the wavelength of the electron beam, $k$ is its radial wave vector, $\theta$ is the transformed azimuthal coordinate and the other terms describe different aberrations. The first two terms in Eq.~(\ref{eq:eq1}) represent the well-known rotationally symmetric aberrations defocus $\Delta f$ and spherical aberration $C_s$. The remaining terms correspond to axial coma (parameterized by $\text{B}_2$ and $\theta_B$), two-fold astigmatism (parameterized by $\text{A}_2$ and $\theta_{2A}$) and three-fold astigmatism (parameterized by $\text{A}_3$ and $\theta_{3A}$), while higher-order aberrations are neglected here. In a STEM that has no multipolar $C_s$ corrector, an appropriate use of probe tilt and quadrupole stigmators allows coma and $\text{A}_2$ to be compensated. Therefore, when $C_s$ is corrected completely, $\text{A}_3$ becomes the first resolution-limiting aberration.
In order to achieve this situation, we aim to impart a phase with a negative value of $C_s$ to the electron beam, while using the defocus $\Delta f$ as an optimization parameter. We therefore work with a nanofabricated hologram, whose thickness is determined by the phase given by the expression
\begin{align}
	\label{eq;eq1}
	\varphi(\rho,\phi) = \varphi_0 F \left[ Q\rho\sin{(\phi)} - \frac{2\pi}{\lambda}\left(\frac{1}{4}C_s \rho^4 - \frac{1}{2}\Delta f \rho^2\right)\right],
\end{align}
where $\rho$ and $\phi$ are polar coordinates in the plane of the hologram and the $Q\rho \sin{(\phi)}$ term enables the use of a probe tilt, which is necessary to produce the main grating frequency. $\varphi_0$ is the modulation depth parameter of the hologram and $F(x)$ is a function that describes the groove profile of the hologram that is required for it to impart a phase of $x$ to the beam. This parameter was modified in our experiment in order to improve the efficiency of the device.
The polar coordinates are calibrated in angle according to the measured convergence of the beam. We aimed for a convergence semi-angle of 12~mrad and assumed a $C_s$ value of $2.7$ mm and a wavelength $\lambda$ of 1.97~pm, which are consistent with the standard conditions of the microscope that was used for the experiment (an FEI TITAN G2 60-300 \cite{boothroyd2016fei}, operated at 300~keV and equipped with a C-twin objective lens). The choice of 12~mrad is close to the maximum value that is allowed in the presence of three-fold astigmatism (whose $\text{A}_3$ parameter is typically close to 1 \textmu m) and other aberrations. Although we deliberately worked with $\Delta f \neq 0$ holograms in some tests, the primary experiments were performed with holograms for which $\Delta f = 0$. A $\Delta f \neq 0$ parameter can be used to extend the region over which the imparted phase varies slowly, which can facilitate fabrication of the hologram.
The hologram produces many diffraction orders, which are associated with different phase modulations. In particular, the $0^{\text{th}}$ diffraction order experiences no phase modulation, while the $n^\text{th}$ diffraction order acquires an ``artificial'' spherical aberration phase of $\varphi = - n \tfrac{2\pi}{\lambda}\left(\tfrac{1}{4}C_s \rho^4\right)$. Only the $1^\text{st}$ diffraction order ($n=1$) acquires a phase profile that exactly compensates for the spherical aberration of the microscope. 
Several additional experimental factors can make the implementation of such a phase profile more difficult. Since we are using a relatively large aperture, the compensation should be as precise as possible, resulting in the need for very precise hologram fabrication. In addition to this requirement, it is necessary to spatially separate each diffraction order from every other one.

The holograms that we used in our experiments were designed using the software STEM CELL~\cite{grillo2013stem_cell} and fabricated using electron beam lithography, according to a similar protocol to that introduced in our previous work~\cite{mafakheri2017realization}. We covered a circular area with a diameter of 80~\textmu m surrounded by a region of SiN covered with Au to produce an effective diaphragm. The resulting membrane is described in Fig.~\ref{fig:fig1}(b),(c). The hologram has $4\text{k}~\times~4\text{k}$ pixels and a rectangular groove profile. This approach results in a piecewise correction of the phase, which becomes smoother at larger radial distances and could be improved in the future by using a focused ion beam (FIB) fabrication process to provide a sinusoidal groove profile.

Just as for in-line nanofabricated holograms, piecewise defined correction can cause spurious intensity components in the probe (\emph{i.e.}, delocalization). Here, we concentrate primarily on the intensity of the contrast transfer function (CTF) in the frequency domain, which should not be affected significantly by this problem. The CTF can be calculated from the transverse phase distribution of the beam in the aperture (\emph{i.e.}, $A(k)=\exp(i(\chi +\varphi))$ for $k$ within the aperture and 0 elsewhere) by using the autocorrelation (here indicated by $\otimes$)~\cite{silcox1992resolution,nellist1998subangstrom}
\begin{align}
	{\rm{CTF}}(k)=A(k)^*\otimes A(k),
\end{align}
which is not affected by phase discontinuities. However, sudden phase jumps may create local intensity modulations in the function $A(k)$. These modulations are visible when the aperture is imaged in ``dark field'' conditions and only one of the beams diffracted by the hologram is selected by the C3 aperture (see Fig.~\ref{fig:fig1}(d)). They were also used as markers to achieve better alignment of the hologram on the beam's propagation axis. 

\section{Results and Discussion}

\begin{figure*}[t]
	\begin{center}
	\includegraphics[width=0.9\linewidth]{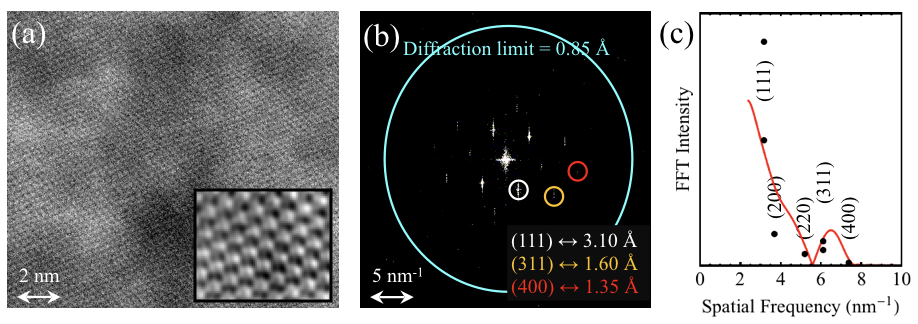}
	\caption[]{a)~Raw STEM image recoded using a single spot $n=1$ (main image) and a filtered image of the lattice (inset). b)~Fourier transform of the experimental STEM image, showing  a faint but visible spot corresponding to the (4,0,0) periodicity in Si. c)~Intensities of the peaks in the Fourier transform plotted as a function of spatial frequency. An example of a simulated CTF with a ``similar'' trend is also shown. We did not try to fit the  (2,0,0) intensity, as it should be almost completely inhibited.}
	\label{fig:fig2}
	\end{center}
\end{figure*}

In order to demonstrate the imaging properties of our setup, we used a test material consisting of a FIB-prepared lamella of Si oriented along [110].
Figure~\ref{fig:fig2}(a) shows an atomic-resolution STEM annular dark-field (ADF) image recorded using a single diffraction probe, which was scanned across the sample following phase modulation by the hologram. For our STEM experiments, the detection semi-angle of the annular detector was set to 24-145 mrad. The inset shows a higher magnification STEM image, displayed following the use of a filtering procedure that is described elsewhere~\cite{boothroyd2016fei}. In the Fourier transform of the STEM image, which is shown in Fig.~\ref{fig:fig2}(b), periodicities as small as 1.36~\AA~, which correspond to the Si (4,0,0) spacing, are present. The visibility of these fringes, which cannot be detected under standard working conditions for which the Scherzer resolution is approximately 1.9\AA, can be considered as a first benchmark. 
Figure~\ref{fig:fig2}(c) shows the effective intensity of these frequency components, plotted alongside a simulated CTF that was obtained by assuming the formation of a linear image $I(k)=\text{CTF}(k)O(k)$ and a simplified object function
\begin{align}
	O(k) = \sum_i \delta(k-k_i)\exp{\left(-\frac{k^2}{\sigma^2}\right)}
\end{align}
with $\sigma=9$ $\text{nm}^{\text{-1}}$. A more detailed description of the function $O(k)$ is discussed in ref.~\cite{grillo2006novel} and references therein.
Although a full dynamical simulation would be more representative of the data in Fig.~\ref{fig:fig2}(c), it is difficult to perform such a calculation without having further physical details about the probe. Nevertheless, it is likely that the CTF that is consistent with the data is characterized by strong intensity up to 4 ${\text{nm}}^{\text{-1}}$ and a secondary region that peaks at 7 ${\text{nm}}^{\text{-1}}$.
It should be noted that the CTF that is shown in Fig.~\ref{fig:fig2}(c) corresponds to a case where spherical aberration is not completely compensated.
This imperfect compensation could have different origins. Given that we used our condenser system in a rather unconventional fashion, our initial hypothesis was that the large aberrations of the condenser could cause a residual effect on the probe. Such effects would not be corrected. However, as there is a large angular magnification generated by the objective, the maximum angle subtended by the condenser aperture is much smaller than (approximately 0.03 times) that subtended by all of the electron beams. Since the aberration phase is defined by the term $C_s k^4$, even an aberration as large as 1~m cannot have a significant influence on the quality of the beam.

Conversely, we estimated that we had an error of up to 3\% in our angle calibration. This seemingly small uncertainty can have a large effect on the shape of the final wavefront. Again, the quartic $k$-dependence of the aberrated phase results in the fact that the hologram produces a compensating phase that is approximately 12\% smaller or larger than that expected.
We estimated that the latter effect corresponds to a maximum residual aberration of $\pm 0.3$ mm.
Considering all of these factors, the CTF shown in Fig.~\ref{fig:fig2} was chosen to correspond to residual aberrations of $C_s=0.3$ mm and $\Delta f =60$ nm.

Other effects may be associated with residual charging of the hologram~\cite{grillo2013stem_cell}. Nevertheless, our data show that the hologram produces a nearly stationary phase distribution up to 8-9~mrad from its center, confirming that the electron-optical configuration performs correctly up to at least this angle. In order to achieve definitive control over $C_s$ in the future, it will be necessary to account for such additional aberrations using an approach that involves better alignment and calibration of the apparatus. All of these factors can be taken care of in future experiments, allowing for near-perfect aberration correction.

Our results are already significant, given the resolution that is demonstrated. They are not, however, intended to compete with multipolar correction. Instead, they are intended to provide a low cost solution for aberration correction, as well as for the two corrector principles to work together, for example by allowing a lower current to be used in the multipolar lenses of a corrector to provide improved stability of its electronics.
This possibility will be a driving force towards improving the present stability of electron-hologram-based correctors, in order to overcoming that of technologically mature state-of-the-art multipolar correctors.

\section{Conclusion}

We have demonstrated an electron-optical configuration for the holographic correction of spherical aberration in a STEM probe. Our setup is based on a nanofabricated off-axis hologram, which introduces a spherical aberration that is opposite to the nominal aberration. We have achieved single beam scanning of a Si [110] sample with a transfer function that extends up to 0.136~nm. Residual aberrations are attributed to additional aberrations introduced by imprecise calibration of the convergence angle of the microscope and to possible charging of the hologram membrane.

\section*{Funding}
V.G. is grateful for support from the Alexander von Humboldt Foundation. H.L. and E.K. acknowledge the support of the Canada Research Chairs (CRC). R.D.-B. is grateful to the European Research Council for funding under the European Union's Seventh Framework Programme (FP7/2007-2013)/ ERC grant agreement number 320832.

\bibliography{references}

\begin{thebibliography}{26}%
\makeatletter
\providecommand \@ifxundefined [1]{%
 \@ifx{#1\undefined}
}%
\providecommand \@ifnum [1]{%
 \ifnum #1\expandafter \@firstoftwo
 \else \expandafter \@secondoftwo
 \fi
}%
\providecommand \@ifx [1]{%
 \ifx #1\expandafter \@firstoftwo
 \else \expandafter \@secondoftwo
 \fi
}%
\providecommand \natexlab [1]{#1}%
\providecommand \enquote  [1]{``#1''}%
\providecommand \bibnamefont  [1]{#1}%
\providecommand \bibfnamefont [1]{#1}%
\providecommand \citenamefont [1]{#1}%
\providecommand \href@noop [0]{\@secondoftwo}%
\providecommand \href [0]{\begingroup \@sanitize@url \@href}%
\providecommand \@href[1]{\@@startlink{#1}\@@href}%
\providecommand \@@href[1]{\endgroup#1\@@endlink}%
\providecommand \@sanitize@url [0]{\catcode `\\12\catcode `\$12\catcode
  `\&12\catcode `\#12\catcode `\^12\catcode `\_12\catcode `\%12\relax}%
\providecommand \@@startlink[1]{}%
\providecommand \@@endlink[0]{}%
\providecommand \url  [0]{\begingroup\@sanitize@url \@url }%
\providecommand \@url [1]{\endgroup\@href {#1}{\urlprefix }}%
\providecommand \urlprefix  [0]{URL }%
\providecommand \Eprint [0]{\href }%
\providecommand \doibase [0]{http://dx.doi.org/}%
\providecommand \selectlanguage [0]{\@gobble}%
\providecommand \bibinfo  [0]{\@secondoftwo}%
\providecommand \bibfield  [0]{\@secondoftwo}%
\providecommand \translation [1]{[#1]}%
\providecommand \BibitemOpen [0]{}%
\providecommand \bibitemStop [0]{}%
\providecommand \bibitemNoStop [0]{.\EOS\space}%
\providecommand \EOS [0]{\spacefactor3000\relax}%
\providecommand \BibitemShut  [1]{\csname bibitem#1\endcsname}%
\let\auto@bib@innerbib\@empty
\bibitem [{\citenamefont {Haider}\ \emph {et~al.}(1998)\citenamefont {Haider},
  \citenamefont {Uhlemann}, \citenamefont {Schwan}, \citenamefont {Rose},
  \citenamefont {Kabius},\ and\ \citenamefont {Urban}}]{haider1998electron}%
  \BibitemOpen
  \bibfield  {author} {\bibinfo {author} {\bibfnamefont {M.}~\bibnamefont
  {Haider}}, \bibinfo {author} {\bibfnamefont {S.}~\bibnamefont {Uhlemann}},
  \bibinfo {author} {\bibfnamefont {E.}~\bibnamefont {Schwan}}, \bibinfo
  {author} {\bibfnamefont {H.}~\bibnamefont {Rose}}, \bibinfo {author}
  {\bibfnamefont {B.}~\bibnamefont {Kabius}}, \ and\ \bibinfo {author}
  {\bibfnamefont {K.}~\bibnamefont {Urban}},\ }\href@noop {} {\bibfield
  {journal} {\bibinfo  {journal} {Nature}\ }\textbf {\bibinfo {volume} {392}},\
  \bibinfo {pages} {768} (\bibinfo {year} {1998})}\BibitemShut {NoStop}%
\bibitem [{\citenamefont {Krivanek}\ \emph {et~al.}(1999)\citenamefont
  {Krivanek}, \citenamefont {Dellby},\ and\ \citenamefont
  {Lupini}}]{krivanek1999towards}%
  \BibitemOpen
  \bibfield  {author} {\bibinfo {author} {\bibfnamefont {O.}~\bibnamefont
  {Krivanek}}, \bibinfo {author} {\bibfnamefont {N.}~\bibnamefont {Dellby}}, \
  and\ \bibinfo {author} {\bibfnamefont {A.}~\bibnamefont {Lupini}},\
  }\href@noop {} {\bibfield  {journal} {\bibinfo  {journal} {Ultramicroscopy}\
  }\textbf {\bibinfo {volume} {78}},\ \bibinfo {pages} {1} (\bibinfo {year}
  {1999})}\BibitemShut {NoStop}%
\bibitem [{\citenamefont {Hawkes}(2009)}]{hawkes2009aberration}%
  \BibitemOpen
  \bibfield  {author} {\bibinfo {author} {\bibfnamefont {P.}~\bibnamefont
  {Hawkes}},\ }\href@noop {} {\bibfield  {journal} {\bibinfo  {journal}
  {Philosophical Transactions of the Royal Society of London A: Mathematical,
  Physical and Engineering Sciences}\ }\textbf {\bibinfo {volume} {367}},\
  \bibinfo {pages} {3637} (\bibinfo {year} {2009})}\BibitemShut {NoStop}%
\bibitem [{\citenamefont {Scherzer}(1936)}]{scherzer1936einige}%
  \BibitemOpen
  \bibfield  {author} {\bibinfo {author} {\bibfnamefont {O.}~\bibnamefont
  {Scherzer}},\ }\href@noop {} {\bibfield  {journal} {\bibinfo  {journal}
  {Zeitschrift f{\"u}r Physik A Hadrons and Nuclei}\ }\textbf {\bibinfo
  {volume} {101}},\ \bibinfo {pages} {593} (\bibinfo {year}
  {1936})}\BibitemShut {NoStop}%
\bibitem [{\citenamefont {Allen}\ \emph {et~al.}(1990)\citenamefont {Allen},
  \citenamefont {Angel}, \citenamefont {Mangus}, \citenamefont {Rodney},
  \citenamefont {Shannon},\ and\ \citenamefont {Spoelhof}}]{allen1990hubble}%
  \BibitemOpen
  \bibfield  {author} {\bibinfo {author} {\bibfnamefont {L.}~\bibnamefont
  {Allen}}, \bibinfo {author} {\bibfnamefont {R.}~\bibnamefont {Angel}},
  \bibinfo {author} {\bibfnamefont {J.~D.}\ \bibnamefont {Mangus}}, \bibinfo
  {author} {\bibfnamefont {G.~A.}\ \bibnamefont {Rodney}}, \bibinfo {author}
  {\bibfnamefont {R.~R.}\ \bibnamefont {Shannon}}, \ and\ \bibinfo {author}
  {\bibfnamefont {C.~P.}\ \bibnamefont {Spoelhof}},\ }\href@noop {} {\bibfield
  {journal} {\bibinfo  {journal} {NASA Report}\ } (\bibinfo {year}
  {1990})}\BibitemShut {NoStop}%
\bibitem [{\citenamefont {Uchida}\ and\ \citenamefont
  {Tonomura}(2010)}]{uchida2010generation}%
  \BibitemOpen
  \bibfield  {author} {\bibinfo {author} {\bibfnamefont {M.}~\bibnamefont
  {Uchida}}\ and\ \bibinfo {author} {\bibfnamefont {A.}~\bibnamefont
  {Tonomura}},\ }\href@noop {} {\bibfield  {journal} {\bibinfo  {journal}
  {nature}\ }\textbf {\bibinfo {volume} {464}},\ \bibinfo {pages} {737}
  (\bibinfo {year} {2010})}\BibitemShut {NoStop}%
\bibitem [{\citenamefont {McMorran}\ \emph {et~al.}(2011)\citenamefont
  {McMorran}, \citenamefont {Agrawal}, \citenamefont {Anderson}, \citenamefont
  {Herzing}, \citenamefont {Lezec}, \citenamefont {McClelland},\ and\
  \citenamefont {Unguris}}]{mcmorran2011electron}%
  \BibitemOpen
  \bibfield  {author} {\bibinfo {author} {\bibfnamefont {B.~J.}\ \bibnamefont
  {McMorran}}, \bibinfo {author} {\bibfnamefont {A.}~\bibnamefont {Agrawal}},
  \bibinfo {author} {\bibfnamefont {I.~M.}\ \bibnamefont {Anderson}}, \bibinfo
  {author} {\bibfnamefont {A.~A.}\ \bibnamefont {Herzing}}, \bibinfo {author}
  {\bibfnamefont {H.~J.}\ \bibnamefont {Lezec}}, \bibinfo {author}
  {\bibfnamefont {J.~J.}\ \bibnamefont {McClelland}}, \ and\ \bibinfo {author}
  {\bibfnamefont {J.}~\bibnamefont {Unguris}},\ }\href@noop {} {\bibfield
  {journal} {\bibinfo  {journal} {science}\ }\textbf {\bibinfo {volume}
  {331}},\ \bibinfo {pages} {192} (\bibinfo {year} {2011})}\BibitemShut
  {NoStop}%
\bibitem [{\citenamefont {Grillo}\ \emph
  {et~al.}(2014{\natexlab{a}})\citenamefont {Grillo}, \citenamefont
  {Carlo~Gazzadi}, \citenamefont {Karimi}, \citenamefont {Mafakheri},
  \citenamefont {Boyd},\ and\ \citenamefont {Frabboni}}]{grillo2014highly}%
  \BibitemOpen
  \bibfield  {author} {\bibinfo {author} {\bibfnamefont {V.}~\bibnamefont
  {Grillo}}, \bibinfo {author} {\bibfnamefont {G.}~\bibnamefont
  {Carlo~Gazzadi}}, \bibinfo {author} {\bibfnamefont {E.}~\bibnamefont
  {Karimi}}, \bibinfo {author} {\bibfnamefont {E.}~\bibnamefont {Mafakheri}},
  \bibinfo {author} {\bibfnamefont {R.~W.}\ \bibnamefont {Boyd}}, \ and\
  \bibinfo {author} {\bibfnamefont {S.}~\bibnamefont {Frabboni}},\ }\href@noop
  {} {\bibfield  {journal} {\bibinfo  {journal} {Applied Physics Letters}\
  }\textbf {\bibinfo {volume} {104}},\ \bibinfo {pages} {043109} (\bibinfo
  {year} {2014}{\natexlab{a}})}\BibitemShut {NoStop}%
\bibitem [{\citenamefont {Verbeeck}\ \emph {et~al.}(2010)\citenamefont
  {Verbeeck}, \citenamefont {Tian},\ and\ \citenamefont
  {Schattschneider}}]{verbeeck2010production}%
  \BibitemOpen
  \bibfield  {author} {\bibinfo {author} {\bibfnamefont {J.}~\bibnamefont
  {Verbeeck}}, \bibinfo {author} {\bibfnamefont {H.}~\bibnamefont {Tian}}, \
  and\ \bibinfo {author} {\bibfnamefont {P.}~\bibnamefont {Schattschneider}},\
  }\href@noop {} {\bibfield  {journal} {\bibinfo  {journal} {Nature}\ }\textbf
  {\bibinfo {volume} {467}},\ \bibinfo {pages} {301} (\bibinfo {year}
  {2010})}\BibitemShut {NoStop}%
\bibitem [{\citenamefont {Shiloh}\ \emph {et~al.}(2014)\citenamefont {Shiloh},
  \citenamefont {Lereah}, \citenamefont {Lilach},\ and\ \citenamefont
  {Arie}}]{shiloh2014sculpturing}%
  \BibitemOpen
  \bibfield  {author} {\bibinfo {author} {\bibfnamefont {R.}~\bibnamefont
  {Shiloh}}, \bibinfo {author} {\bibfnamefont {Y.}~\bibnamefont {Lereah}},
  \bibinfo {author} {\bibfnamefont {Y.}~\bibnamefont {Lilach}}, \ and\ \bibinfo
  {author} {\bibfnamefont {A.}~\bibnamefont {Arie}},\ }\href@noop {} {\bibfield
   {journal} {\bibinfo  {journal} {Ultramicroscopy}\ }\textbf {\bibinfo
  {volume} {144}},\ \bibinfo {pages} {26} (\bibinfo {year} {2014})}\BibitemShut
  {NoStop}%
\bibitem [{\citenamefont {Voloch-Bloch}\ \emph {et~al.}(2013)\citenamefont
  {Voloch-Bloch}, \citenamefont {Lereah}, \citenamefont {Lilach}, \citenamefont
  {Gover},\ and\ \citenamefont {Arie}}]{voloch2013generation}%
  \BibitemOpen
  \bibfield  {author} {\bibinfo {author} {\bibfnamefont {N.}~\bibnamefont
  {Voloch-Bloch}}, \bibinfo {author} {\bibfnamefont {Y.}~\bibnamefont
  {Lereah}}, \bibinfo {author} {\bibfnamefont {Y.}~\bibnamefont {Lilach}},
  \bibinfo {author} {\bibfnamefont {A.}~\bibnamefont {Gover}}, \ and\ \bibinfo
  {author} {\bibfnamefont {A.}~\bibnamefont {Arie}},\ }\href@noop {} {\bibfield
   {journal} {\bibinfo  {journal} {Nature}\ }\textbf {\bibinfo {volume}
  {494}},\ \bibinfo {pages} {331} (\bibinfo {year} {2013})}\BibitemShut
  {NoStop}%
\bibitem [{\citenamefont {Grillo}\ \emph
  {et~al.}(2014{\natexlab{b}})\citenamefont {Grillo}, \citenamefont {Karimi},
  \citenamefont {Gazzadi}, \citenamefont {Frabboni}, \citenamefont {Dennis},\
  and\ \citenamefont {Boyd}}]{grillo2014generation}%
  \BibitemOpen
  \bibfield  {author} {\bibinfo {author} {\bibfnamefont {V.}~\bibnamefont
  {Grillo}}, \bibinfo {author} {\bibfnamefont {E.}~\bibnamefont {Karimi}},
  \bibinfo {author} {\bibfnamefont {G.~C.}\ \bibnamefont {Gazzadi}}, \bibinfo
  {author} {\bibfnamefont {S.}~\bibnamefont {Frabboni}}, \bibinfo {author}
  {\bibfnamefont {M.~R.}\ \bibnamefont {Dennis}}, \ and\ \bibinfo {author}
  {\bibfnamefont {R.~W.}\ \bibnamefont {Boyd}},\ }\href@noop {} {\bibfield
  {journal} {\bibinfo  {journal} {Physical Review X}\ }\textbf {\bibinfo
  {volume} {4}},\ \bibinfo {pages} {011013} (\bibinfo {year}
  {2014}{\natexlab{b}})}\BibitemShut {NoStop}%
\bibitem [{\citenamefont {Grillo}\ \emph {et~al.}(2017)\citenamefont {Grillo},
  \citenamefont {Tavabi}, \citenamefont {Venturi}, \citenamefont {Larocque},
  \citenamefont {Balboni}, \citenamefont {Gazzadi}, \citenamefont {Frabboni},
  \citenamefont {Lu}, \citenamefont {Mafakheri}, \citenamefont {Bouchard},
  \citenamefont {Dunin-Borkowski}, \citenamefont {Boyd}, \citenamefont
  {Lavery}, \citenamefont {Padgett},\ and\ \citenamefont
  {Karimi}}]{grillo2017measuring}%
  \BibitemOpen
  \bibfield  {author} {\bibinfo {author} {\bibfnamefont {V.}~\bibnamefont
  {Grillo}}, \bibinfo {author} {\bibfnamefont {A.~H.}\ \bibnamefont {Tavabi}},
  \bibinfo {author} {\bibfnamefont {F.}~\bibnamefont {Venturi}}, \bibinfo
  {author} {\bibfnamefont {H.}~\bibnamefont {Larocque}}, \bibinfo {author}
  {\bibfnamefont {R.}~\bibnamefont {Balboni}}, \bibinfo {author} {\bibfnamefont
  {G.~C.}\ \bibnamefont {Gazzadi}}, \bibinfo {author} {\bibfnamefont
  {S.}~\bibnamefont {Frabboni}}, \bibinfo {author} {\bibfnamefont {P.-H.}\
  \bibnamefont {Lu}}, \bibinfo {author} {\bibfnamefont {E.}~\bibnamefont
  {Mafakheri}}, \bibinfo {author} {\bibfnamefont {F.}~\bibnamefont {Bouchard}},
  \bibinfo {author} {\bibfnamefont {R.~E.}\ \bibnamefont {Dunin-Borkowski}},
  \bibinfo {author} {\bibfnamefont {R.~W.}\ \bibnamefont {Boyd}}, \bibinfo
  {author} {\bibfnamefont {M.~P.~J.}\ \bibnamefont {Lavery}}, \bibinfo {author}
  {\bibfnamefont {M.~J.}\ \bibnamefont {Padgett}}, \ and\ \bibinfo {author}
  {\bibfnamefont {E.}~\bibnamefont {Karimi}},\ }\href@noop {} {\bibfield
  {journal} {\bibinfo  {journal} {Nature Communications}\ }\textbf {\bibinfo
  {volume} {8}},\ \bibinfo {pages} {15536} (\bibinfo {year}
  {2017})}\BibitemShut {NoStop}%
\bibitem [{\citenamefont {Linck}\ \emph {et~al.}(2014)\citenamefont {Linck},
  \citenamefont {McMorran}, \citenamefont {Pierce},\ and\ \citenamefont
  {Ercius}}]{linck2014aberration}%
  \BibitemOpen
  \bibfield  {author} {\bibinfo {author} {\bibfnamefont {M.}~\bibnamefont
  {Linck}}, \bibinfo {author} {\bibfnamefont {B.}~\bibnamefont {McMorran}},
  \bibinfo {author} {\bibfnamefont {J.}~\bibnamefont {Pierce}}, \ and\ \bibinfo
  {author} {\bibfnamefont {P.}~\bibnamefont {Ercius}},\ }\href@noop {}
  {\bibfield  {journal} {\bibinfo  {journal} {Microscopy and Microanalysis}\
  }\textbf {\bibinfo {volume} {20}},\ \bibinfo {pages} {946} (\bibinfo {year}
  {2014})}\BibitemShut {NoStop}%
\bibitem [{\citenamefont {Grillo}\ \emph {et~al.}(2015)\citenamefont {Grillo},
  \citenamefont {Pierce}, \citenamefont {Karimi}, \citenamefont {Harvey},
  \citenamefont {Balboni}, \citenamefont {Gazzadi}, \citenamefont {Mafakheri},
  \citenamefont {Venturi}, \citenamefont {McMorran}, \citenamefont {Frabboni}
  \emph {et~al.}}]{grillo2015structured}%
  \BibitemOpen
  \bibfield  {author} {\bibinfo {author} {\bibfnamefont {V.}~\bibnamefont
  {Grillo}}, \bibinfo {author} {\bibfnamefont {J.~S.}\ \bibnamefont {Pierce}},
  \bibinfo {author} {\bibfnamefont {E.}~\bibnamefont {Karimi}}, \bibinfo
  {author} {\bibfnamefont {T.~R.}\ \bibnamefont {Harvey}}, \bibinfo {author}
  {\bibfnamefont {R.}~\bibnamefont {Balboni}}, \bibinfo {author} {\bibfnamefont
  {G.~C.}\ \bibnamefont {Gazzadi}}, \bibinfo {author} {\bibfnamefont
  {E.}~\bibnamefont {Mafakheri}}, \bibinfo {author} {\bibfnamefont
  {F.}~\bibnamefont {Venturi}}, \bibinfo {author} {\bibfnamefont {B.~J.}\
  \bibnamefont {McMorran}}, \bibinfo {author} {\bibfnamefont {S.}~\bibnamefont
  {Frabboni}},  \emph {et~al.},\ }\href@noop {} {\bibfield  {journal} {\bibinfo
   {journal} {Microscopy and Microanalysis}\ }\textbf {\bibinfo {volume}
  {21}},\ \bibinfo {pages} {25} (\bibinfo {year} {2015})}\BibitemShut {NoStop}%
\bibitem [{\citenamefont {Grillo}\ \emph
  {et~al.}(2014{\natexlab{c}})\citenamefont {Grillo}, \citenamefont {Karimi},
  \citenamefont {Balboni}, \citenamefont {Gazzadi}, \citenamefont {Frabboni},
  \citenamefont {Mafakheri},\ and\ \citenamefont {Boyd}}]{grilloToward2014}%
  \BibitemOpen
  \bibfield  {author} {\bibinfo {author} {\bibfnamefont {V.}~\bibnamefont
  {Grillo}}, \bibinfo {author} {\bibfnamefont {E.}~\bibnamefont {Karimi}},
  \bibinfo {author} {\bibfnamefont {R.}~\bibnamefont {Balboni}}, \bibinfo
  {author} {\bibfnamefont {G.~C.}\ \bibnamefont {Gazzadi}}, \bibinfo {author}
  {\bibfnamefont {S.}~\bibnamefont {Frabboni}}, \bibinfo {author}
  {\bibfnamefont {E.}~\bibnamefont {Mafakheri}}, \ and\ \bibinfo {author}
  {\bibfnamefont {R.~W.}\ \bibnamefont {Boyd}},\ }\href@noop {} {\bibfield
  {journal} {\bibinfo  {journal} {IMC conference}\ }\textbf {\bibinfo {volume}
  {IT-1-P-6140}} (\bibinfo {year} {2014}{\natexlab{c}})}\BibitemShut {NoStop}%
\bibitem [{\citenamefont {Shiloh}\ \emph {et~al.}(2016)\citenamefont {Shiloh},
  \citenamefont {Remez},\ and\ \citenamefont {Arie}}]{shiloh2016prospects}%
  \BibitemOpen
  \bibfield  {author} {\bibinfo {author} {\bibfnamefont {R.}~\bibnamefont
  {Shiloh}}, \bibinfo {author} {\bibfnamefont {R.}~\bibnamefont {Remez}}, \
  and\ \bibinfo {author} {\bibfnamefont {A.}~\bibnamefont {Arie}},\ }\href@noop
  {} {\bibfield  {journal} {\bibinfo  {journal} {Ultramicroscopy}\ }\textbf
  {\bibinfo {volume} {163}},\ \bibinfo {pages} {69} (\bibinfo {year}
  {2016})}\BibitemShut {NoStop}%
\bibitem [{\citenamefont {Krivanek}\ \emph {et~al.}(2014)\citenamefont
  {Krivanek}, \citenamefont {Rusz}, \citenamefont {Idrobo}, \citenamefont
  {Lovejoy},\ and\ \citenamefont {Dellby}}]{krivanek2014toward}%
  \BibitemOpen
  \bibfield  {author} {\bibinfo {author} {\bibfnamefont {O.~L.}\ \bibnamefont
  {Krivanek}}, \bibinfo {author} {\bibfnamefont {J.}~\bibnamefont {Rusz}},
  \bibinfo {author} {\bibfnamefont {J.-C.}\ \bibnamefont {Idrobo}}, \bibinfo
  {author} {\bibfnamefont {T.~J.}\ \bibnamefont {Lovejoy}}, \ and\ \bibinfo
  {author} {\bibfnamefont {N.}~\bibnamefont {Dellby}},\ }\href@noop {}
  {\bibfield  {journal} {\bibinfo  {journal} {Microscopy and Microanalysis}\
  }\textbf {\bibinfo {volume} {20}},\ \bibinfo {pages} {832} (\bibinfo {year}
  {2014})}\BibitemShut {NoStop}%
\bibitem [{\citenamefont {Pohl}\ \emph {et~al.}()\citenamefont {Pohl},
  \citenamefont {Rusz}, \citenamefont {Spiegelberg}, \citenamefont {Schneider},
  \citenamefont {Tiemeijer}, \citenamefont {Nielsch},\ and\ \citenamefont
  {Rellinghaus}}]{pohl2016towards}%
  \BibitemOpen
  \bibfield  {author} {\bibinfo {author} {\bibfnamefont {D.}~\bibnamefont
  {Pohl}}, \bibinfo {author} {\bibfnamefont {J.}~\bibnamefont {Rusz}}, \bibinfo
  {author} {\bibfnamefont {J.}~\bibnamefont {Spiegelberg}}, \bibinfo {author}
  {\bibfnamefont {S.}~\bibnamefont {Schneider}}, \bibinfo {author}
  {\bibfnamefont {P.}~\bibnamefont {Tiemeijer}}, \bibinfo {author}
  {\bibfnamefont {K.}~\bibnamefont {Nielsch}}, \ and\ \bibinfo {author}
  {\bibfnamefont {B.}~\bibnamefont {Rellinghaus}},\ }in\ \href@noop {} {\emph
  {\bibinfo {booktitle} {European Microscopy Congress 2016: Proceedings}}}\
  (\bibinfo {organization} {Wiley Online Library})\BibitemShut {NoStop}%
\bibitem [{\citenamefont {Kirkland}(2010)}]{kirkland2010advanced}%
  \BibitemOpen
  \bibfield  {author} {\bibinfo {author} {\bibfnamefont {E.~J.}\ \bibnamefont
  {Kirkland}},\ }\href@noop {} {\emph {\bibinfo {title} {Advanced computing in
  electron microscopy}}}\ (\bibinfo  {publisher} {Springer Science \&amp;
  Business Media},\ \bibinfo {year} {2010})\BibitemShut {NoStop}%
\bibitem [{\citenamefont {Boothroyd}\ \emph {et~al.}(2016)\citenamefont
  {Boothroyd}, \citenamefont {Kov{\'a}cs},\ and\ \citenamefont
  {Tillmann}}]{boothroyd2016fei}%
  \BibitemOpen
  \bibfield  {author} {\bibinfo {author} {\bibfnamefont {C.}~\bibnamefont
  {Boothroyd}}, \bibinfo {author} {\bibfnamefont {A.}~\bibnamefont
  {Kov{\'a}cs}}, \ and\ \bibinfo {author} {\bibfnamefont {K.}~\bibnamefont
  {Tillmann}},\ }\href@noop {} {\bibfield  {journal} {\bibinfo  {journal}
  {Journal of large-scale research facilities JLSRF}\ }\textbf {\bibinfo
  {volume} {2}},\ \bibinfo {pages} {44} (\bibinfo {year} {2016})}\BibitemShut
  {NoStop}%
\bibitem [{\citenamefont {Grillo}\ and\ \citenamefont
  {Rossi}(2013)}]{grillo2013stem_cell}%
  \BibitemOpen
  \bibfield  {author} {\bibinfo {author} {\bibfnamefont {V.}~\bibnamefont
  {Grillo}}\ and\ \bibinfo {author} {\bibfnamefont {F.}~\bibnamefont {Rossi}},\
  }\href@noop {} {\bibfield  {journal} {\bibinfo  {journal} {Ultramicroscopy}\
  }\textbf {\bibinfo {volume} {125}},\ \bibinfo {pages} {112} (\bibinfo {year}
  {2013})}\BibitemShut {NoStop}%
\bibitem [{\citenamefont {Mafakheri}\ \emph {et~al.}(2017)\citenamefont
  {Mafakheri}, \citenamefont {Tavabi}, \citenamefont {Lu}, \citenamefont
  {Balboni}, \citenamefont {Venturi}, \citenamefont {Menozzi}, \citenamefont
  {Gazzadi}, \citenamefont {Frabboni}, \citenamefont {Sit}, \citenamefont
  {Dunin-Borkowski} \emph {et~al.}}]{mafakheri2017realization}%
  \BibitemOpen
  \bibfield  {author} {\bibinfo {author} {\bibfnamefont {E.}~\bibnamefont
  {Mafakheri}}, \bibinfo {author} {\bibfnamefont {A.}~\bibnamefont {Tavabi}},
  \bibinfo {author} {\bibfnamefont {P.-H.}\ \bibnamefont {Lu}}, \bibinfo
  {author} {\bibfnamefont {R.}~\bibnamefont {Balboni}}, \bibinfo {author}
  {\bibfnamefont {F.}~\bibnamefont {Venturi}}, \bibinfo {author} {\bibfnamefont
  {C.}~\bibnamefont {Menozzi}}, \bibinfo {author} {\bibfnamefont
  {G.}~\bibnamefont {Gazzadi}}, \bibinfo {author} {\bibfnamefont
  {S.}~\bibnamefont {Frabboni}}, \bibinfo {author} {\bibfnamefont
  {A.}~\bibnamefont {Sit}}, \bibinfo {author} {\bibfnamefont {R.}~\bibnamefont
  {Dunin-Borkowski}},  \emph {et~al.},\ }\href@noop {} {\bibfield  {journal}
  {\bibinfo  {journal} {Applied Physics Letters}\ }\textbf {\bibinfo {volume}
  {110}},\ \bibinfo {pages} {093113} (\bibinfo {year} {2017})}\BibitemShut
  {NoStop}%
\bibitem [{\citenamefont {Silcox}\ \emph {et~al.}(1992)\citenamefont {Silcox},
  \citenamefont {Xu},\ and\ \citenamefont {Loane}}]{silcox1992resolution}%
  \BibitemOpen
  \bibfield  {author} {\bibinfo {author} {\bibfnamefont {J.}~\bibnamefont
  {Silcox}}, \bibinfo {author} {\bibfnamefont {P.}~\bibnamefont {Xu}}, \ and\
  \bibinfo {author} {\bibfnamefont {R.~F.}\ \bibnamefont {Loane}},\ }\href@noop
  {} {\bibfield  {journal} {\bibinfo  {journal} {Ultramicroscopy}\ }\textbf
  {\bibinfo {volume} {47}},\ \bibinfo {pages} {173} (\bibinfo {year}
  {1992})}\BibitemShut {NoStop}%
\bibitem [{\citenamefont {Nellist}\ and\ \citenamefont
  {Pennycook}(1998)}]{nellist1998subangstrom}%
  \BibitemOpen
  \bibfield  {author} {\bibinfo {author} {\bibfnamefont {P.}~\bibnamefont
  {Nellist}}\ and\ \bibinfo {author} {\bibfnamefont {S.}~\bibnamefont
  {Pennycook}},\ }\href@noop {} {\bibfield  {journal} {\bibinfo  {journal}
  {Physical Review Letters}\ }\textbf {\bibinfo {volume} {81}},\ \bibinfo
  {pages} {4156} (\bibinfo {year} {1998})}\BibitemShut {NoStop}%
\bibitem [{\citenamefont {Grillo}\ and\ \citenamefont
  {Carlino}(2006)}]{grillo2006novel}%
  \BibitemOpen
  \bibfield  {author} {\bibinfo {author} {\bibfnamefont {V.}~\bibnamefont
  {Grillo}}\ and\ \bibinfo {author} {\bibfnamefont {E.}~\bibnamefont
  {Carlino}},\ }\href@noop {} {\bibfield  {journal} {\bibinfo  {journal}
  {Ultramicroscopy}\ }\textbf {\bibinfo {volume} {106}},\ \bibinfo {pages}
  {603} (\bibinfo {year} {2006})}\BibitemShut {NoStop}%
\end{thebibliography}%

\end{document}